\documentclass[%
 reprint,
 amsmath,amssymb,
 aps, physrev,
showkeys
]{revtex4-2}

\usepackage{graphicx}
\usepackage{subfigure}
\usepackage{dcolumn}
\usepackage{bm}

\usepackage{color}
\graphicspath{{/figs/}}

\begin{document}

\preprint{APS/123-QED}

\title{\textbf{Twist dominates bending in the liquid crystal organization of bacteriophage DNA} 
}%

\author{Pei Liu}
\affiliation{%
 Department of Mathematics and Systems Engineering, Florida Institute of Technology, Melbourne, FL, 32901
}%
\author{Tamara Christiani}%
\affiliation{
 Integrative Genetics and Genomics Graduate Group, University of California, Davis, CA 95616
}
\thanks{ P.L. (Pei Liu) and T.C.(Tamara Christiani) contributed equally to this work}
\author{Zhijie Wang}
 \affiliation{
 Graduate Program in Applied Mathematics, University of California, Davis, CA 95616
}
\author{Fei Guo}
 \affiliation{
 Biological Electron Microscopy Facility, University of California, Davis, CA 95616
}
\author{Mariel Vazquez}
  \affiliation{
Department of Mathematics, University of California, Davis, CA 95616
}
 \affiliation{
 Department of Microbiology and Molecular Genetics, University of California, Davis, CA 95616
}
\author{Carme Calderer} 
 \affiliation{
 School of Mathematics, University of Minnesota, Minneapolis, MN 55442
}
\author{Javier Arsuaga}

\affiliation{Department of Molecular and Cellular Biology, University of California, Davis, CA 95616}

 \affiliation{
Department of Mathematics, University of California, Davis, CA 95616
}
\email{Corresponding author: jarsuaga@ucdavis.edu}

\date{\today}

\begin{abstract}
DNA frequently adopts liquid-crystalline conformations in both cells and viruses. The Oseen--Frank framework provides a powerful continuum description of these phases through three elastic moduli: splay ($K_1$), twist or cholesteric ($K_2$), and bending ($K_3$). While $K_1$ is typically assumed to dominate, the relative magnitude of $K_2$ and $K_3$ in confined DNA remains poorly understood. Here, we combine cryo-electron microscopy, liquid-crystal modeling, and knot theory to quantify this relationship in bacteriophage P4, whose genome is partially organized in a spool-like liquid-crystalline phase. We first show experimentally that the ordered DNA occupies three concentric layers within the capsid. We then formulate an Oseen--Frank model for this geometry and use it, together with the measured layer radii, to estimate the elastic ratio $\alpha = K_3/K_2$. We find $\alpha \approx 0.0064$, indicating that twist elasticity overwhelmingly dominates bending. To validate this result, we perform Langevin dynamics simulations of DNA trajectories and classify the resulting knots. The predicted knot distribution agrees with experimental data from P4, demonstrating consistency between elasticity, topology, and observed genome organization.
\end{abstract}

\maketitle



{\it Introduction ---} Double stranded (ds) DNA organizes 
into a liquid crystal state when it reaches densities comparable to those found in many living systems \cite{livolant1984cholesteric,livolant1978positive,reich1994liquid,rill1989electron} and viruses \cite{Leforestier1993,livolant1984cholesteric, Pelta1996, strzelecka1988multiple},
such as herpesviruses \cite{booy1991liquid} and bacteriophages \cite{lepault1987,Leforestier2009,Leforestier2010}. Owing to their structural simplicity, bacteriophages offer a compelling model for probing DNA liquid crystalline phases \cite{Leforestier1993,Leforestier2009,lepault1987}, as a single naked DNA molecule is densely confined within an icosahedral protein capsid \cite{Comolli2008,Jiang2006,johnson2007dna}.

Average cryo-electron microscopy (cryo-EM) density maps of bacteriophage particles reveal two patterns of genome organization  \cite{Cerritelli1997,Comolli2008,Earnshaw1980,Lander2013,liu2005nature}. Near the inner surface of the capsid, the phage DNA is arranged  in concentric layers that spool around the central axis of the capsid, with an  average inter-helix spacing of  approximately $2.5 \ nm$ \cite{Earnshaw1980,Cerritelli1997}. This spooled region is estimated to contain at least 50\% of the viral genome \cite{Comolli2008}. A  second  organizational pattern emerges near the center of the capsid, where the DNA appears  disordered or isotropic. This region lacks a well-defined structure and may vary between viral particles, resulting in a diffuse signal in cryo-EM reconstructions.

 The Oseen--Frank free energy density of the nematic phase has been a powerful
 workhorse in the equilibrium studies of liquid crystals since the  1930's \cite{oseen1933theory}. Formulated as a quadratic function of the unit director field $\vec{n}$,  the average  direction of alignment of rigid molecular groups, it involves three positive coefficients $K_1, K_2$ and $K_3$, the splay, twist and bend elastic Frank constants, representing the energy contributions corresponding to such deformations. Originally developed for small molecule liquid crystal materials, the Oseen--Frank energy has been further adapted to polymeric materials --- long chains  with embedded  rigid units --- such as semiflexible polymers and, notably, DNA \cite{vroege1988induced, hiltner2018equilibrium,klug2003director}. In this context, minimizing the total energy of DNA leads to an optimal unit vector field, $\vec{n}$, that locally describes the preferred orientation of the polymer segments. Reconstructing the linear trajectory of the axis of the DNA molecule requires further integration along the field $\vec{n}$ \cite{klug2003director,liu2022helical}.  

One limitation of the Oseen--Frank energy formulation lies in  the difficulty of accurately  measuring   the Frank constants, particularly in condensed phases of semiflexible polymers such as DNA. Analytical expressions for the bending modulus $K_3$, based on the polymer’s persistence length, density, and surrounding ionic concentrations, have been proposed \cite{brunet2015dependence}. Both theoretical studies on polymers \cite{vroege1988induced} and experimental studies on DNA \cite{lucchetti2020elasticity} indicate that $K_1$ is significantly larger than both $K_2$ and $K_3$. 
This difference in values aligns with the characteristic layered or spool-like organization observed in condensed DNA structures \cite{hud2005,Cerritelli1997}. 
These studies further report that $K_3>K_2$. However, the consistent arrangement of DNA into quasi-parallel spooling layers across a broad range of ionic conditions appears to favor the opposite inequality \cite{Lander2013,qiu2011salt}.

A second limitation of the Oseen--Frank formulation is its assignment of infinite energy to liquid crystal defects --- such as point, line, and planar singularities --- which are, in fact, ubiquitous in liquid crystal systems. In the context of DNA studies, this presents a significant drawback: the Oseen-Frank energy, along with its reconstruction of the molecule’s linear trajectory, fail to capture topological features such as local crossings of DNA fibers and knots, which are commonly observed in DNA condensates \cite{Arsuaga2002b,hud2005,liu1981knotted}.

In this study, we combine experimental imaging with analytical and numerical methods to estimate the  ratio of elastic constants $\alpha=\frac{K_3}{K_2}$ for DNA within bacteriophage capsids. 

{\it Experimental setup ---}We amplified and purified mature bacteriophage P4 vir1 del22, a 1.7 kb deletion mutant of bacteriophage P4 vir1  \cite{raimondi1985analysis}, using strain C-1895 in liquid culture, as described in \cite{Isaksen1999}.  For clarity and  brevity, we will refer to P4 vir1 del22 as P4 throughout the text.
\begin{figure}[htpb]
   \centering
\includegraphics[scale=0.32]{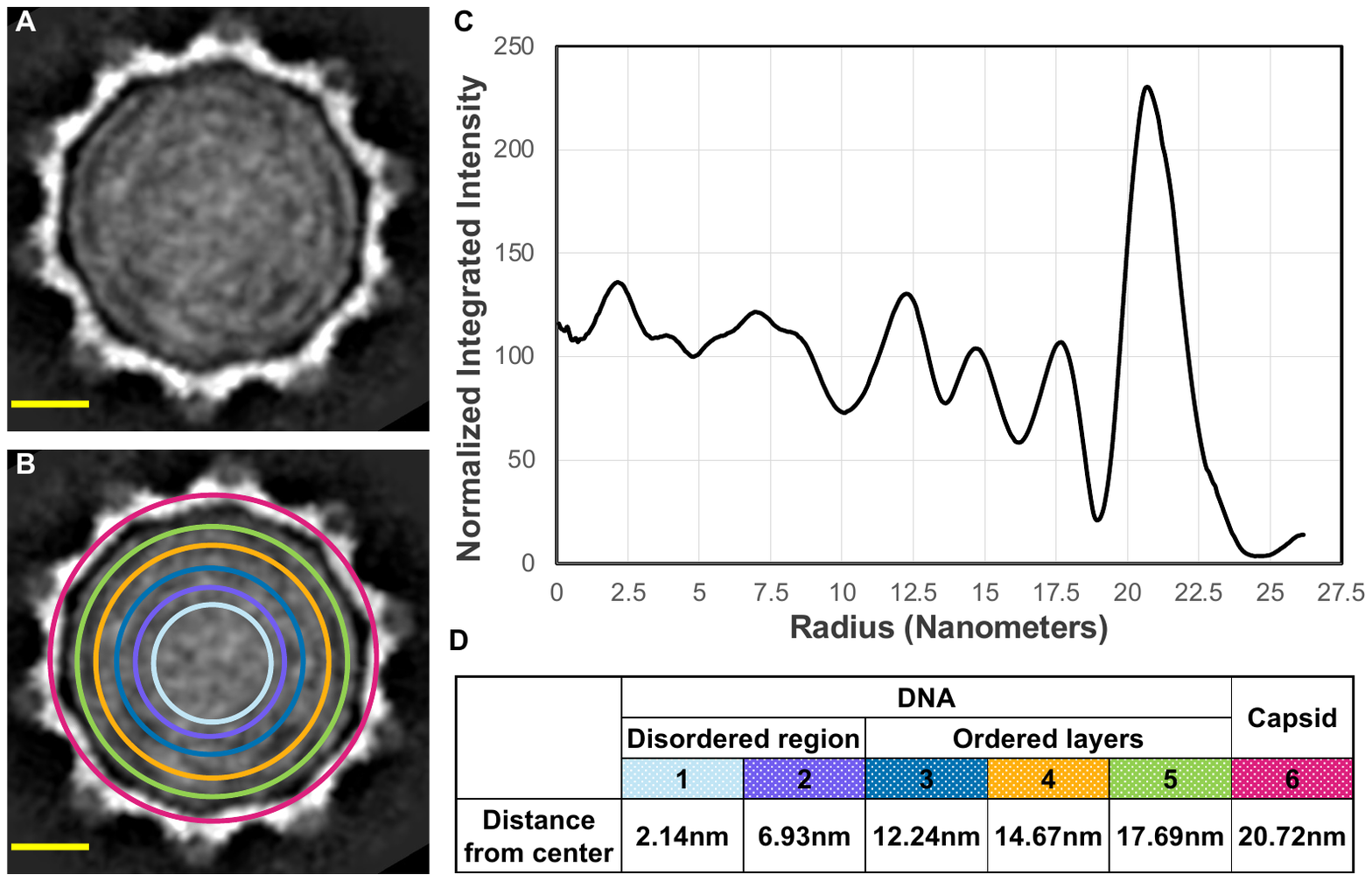} 
\caption{Concentric layer architecture observed in P4 bacteriophages: A) Central slice along a 5 fold symmetry axis of the 3D cryoEM reconstruction of P4 nucleocapsid B) Same as A) with concentric DNA and capsid layers identified  C) Pixel normalized radial intensity profile of image shown in A). D) Chart radial values for concentric layers. Only layers 3 to 6 are consistent across projections.}\label{fig:P4_layers}
\end{figure}

Next, we collected over 9,000 images of bacteriophage P4.  We used CryoSPARC \cite{punjani2017cryosparc}, ChimeraX \cite{pettersen2021ucsf}, and Fiji\cite{schindelin2012fiji}  to identify phage particles, compute their average electron density with a resolution of 13.14 \AA, and extract the number of layers associated with the spooling region of P4 DNA. Results are shown in Fig. \ref{fig:P4_layers}. Panel A shows the central slice orthogonal to the five fold axis of symmetry along the axis of the tail. Panel B highlights the estimated location of the DNA layers and the capsid layer, as indicated by the highest local pixel intensity.  Panel C displays the corresponding density graph, with the maxima  corresponding to the regions of highest pixel intensity.  Panel D shows a table with the radial distances for each detected layer. The outer most layer, with a center line at a radius of $20.72 \ nm$ (indicated by the red contour in panel B), corresponds to the protein capsid \cite{dokland1992image}. Inside the capsid, three concentric DNA layers are clearly visible: an outer layer (green contour in panel B) with a  radius of  $17.69 \ nm$,  a second concentric layer (yellow contour in panel B) with a  radius of $14.67 \ nm$, and a third layer (dark blue contour in panel B) with a radius of $12.24 \ nm$. The figure shows additional layers, but only the outer three layers were consistently observed  across all axes of symmetry (See SI). Based on these observations, we conclude that the spooling region consists of three layers and it is confined to the region delimited by radii $10.0 \ nm$ and $20.7 \ nm$.  The distance between the centerline of the capsid and  the outermost DNA layer is $3.0 \ nm$, confirming the repulsive nature of the capsid \cite{Lander2013,coshic2024structure}. The distances between the three consecutive DNA layers, and between the third DNA layer and the boundary of the disordered region are consistent with data observed in other deletion mutant viruses \cite{Lander2013,qiu2011salt} and in agreement with molecular mechanics predictions on bacteriophage P4\cite{Petrov2011}. The radius of the disordered phase (local minimum at $10.0 \ nm$ in Panel D) also agrees with the value of $9.48 \ nm$ predicted by a chromonic liquid crystal model \cite{walker2020fine}. Based on these findings, we  conclude that bacteriophage P4 contains three layers of spool-like ordered DNA  near the capsid, along with  a disordered region whose radius is approximately half of the capsid radius. 

{\it Oseen--Frank Theory ---
} 
We denote by $\Omega$ the region confined by the capsid that contains the spooling model of DNA, 
 excluding the polar caps and the disordered core of the capsid. 
In cylindrical coordinates: 
 $$\Omega=\{(r, \theta, z): R_1 <  r< R_2,  0\leq \theta\leq 2\pi,  0\leq z\leq H\}.$$  
In keeping with the observation 
that the DNA in the region $\Omega$ is organized as a chromonic liquid crystal with density $\rho(\vec{x})$ \cite{hiltner2021chromonic}, we represent the DNA molecule by a unit vector field-line pair $(\vec{n}, \mathcal L)$, where  $\vec{n}(\vec{x})$ minimizes the total Oseen--Frank energy,
\begin{equation}
    E_{OF} = \int_\Omega \rho \left[ K_3 |\vec{n} \times \nabla \times \vec{n}|^2 + K_2 (\vec{n} \cdot \nabla \times \vec{n})^2\right] d\vec{x}. \label{total energy with inner core}
\end{equation}
The energy in equation \eqref{total energy with inner core} is minimized,  subject to the constraint $\nabla\cdot\vec{n}=0$, which   arises in the limiting case $K_1\to\infty$. The boundary conditions will be specified later.
The line $\mathcal L$, representing the trajectory of the center axis of the DNA molecule, is described  by a smooth curve $\vec{r}(s)$ that  solves the following initial value problem 
\begin{equation}
    \frac{d \vec{r}}{ds} = \vec{n}(s), \,\vec{r}(0) =\vec{r}_0.
\end{equation}
Here $s$ denotes the arc length along the curve, and $\vec{r}_0$ specifies the location of one end of the DNA.

In this study we consider the model presented in \cite{liu2025knotted}. We focus on the interplay between bending and twist contribution while ignoring the entropic effect. The helical curve of DNA can be parameterized by,
\begin{equation}
    \vec{n}(r,\theta,z) =  \cos\psi \vec{e}_\theta+ \sin\psi\vec{e}_z, \quad \psi=\psi(r, \theta, z),
\end{equation}
where $\{\vec{e}_r, \vec{e}_\theta, \vec{e}_z\}$ is the orthogonal basis of the  cylindrical coordinate system $(r, \theta, z)$, and $\psi \in [0,\pi/2]$  denotes the local helical angle. 
The pitch of a helix is defined as  the vertical distance (along the helical axis)   corresponding to a complete rotation of $2\pi$ radians. 
That is, 
\begin{equation}
P(r,\theta,z) = 2 \pi r \tan \psi. 
\end{equation}
This expression captures how the helix stretches vertically as a function of its local tilt $\psi$. Additionally, we prescribe the distance between neighboring DNA segments to be of the same order of magnitude as the helical pitch, in order to remain consistent with the hexagonal lattice structure characteristic of the spooling configuration. This allows us to postulate that the DNA local density satisfies,

\begin{equation}
    \displaystyle \rho(r,\theta,z) = \frac{C}{r^2 \tan^2 \psi}. \label{3Ddensity}
\end{equation}
The parameter $C= \frac{\eta}{4\pi^2\sqrt{3}}$ 
arises from the hexagonal packing geometry of the spooled DNA lattice, and $\eta= 3{\text{nm}}^{-1} $ denotes the line density of DNA base pairs. For a more detailed mathematical derivation see \cite{liu2022helical,liu2025knotted}.
The minimizer of the energy functional in \eqref{total energy with inner core} satisfies the associated  Euler--Lagrange equation, expressed  in terms of the transformed variables,
$y=-\log\frac{r}{R_2}, \, u = -\ln \sin \psi$ gives:
\begin{equation}
     \big(e^{2y}u_y\big)_y- 2e^{2y}\big((\alpha-1)e^{-4u} + \frac{\alpha}{2}(e^{2u}-3 e^{-2u}) + 1\big)=0, \, \label{hy-eqn}
\end{equation} 
$y\in [0, \log\frac{R_2}{R_1}]$. Furthermore, we impose Dirichlet boundary conditions on both ends, 
\begin{equation}
 u(\log\frac{R_2}{R_1})=M_1. \,\, \, u(0)=M_2,  \label{BC}
 \end{equation}
 with $M_1$ and $M_2$ defined as follows.  At the outer boundary $r=R_2$, we assume a vanishing helical pitch $p = \epsilon \to 0$,  which reflects a strong anchoring to a concentric  spooling configuration, characterized by a small   helical angle $\psi(R_2) = \arctan \frac{\epsilon}{2\pi R_2}$  \cite{liu2022helical}. Consequently, the boundary value   $M_2 = |\ln \sin \psi(R_2)|$  is  large since  $|\psi(R_2)|$ is small. 

 At the inner boundary $r=R_1$, where the disordered core begins, we assume the helical pitch equals the radial value $R_1$. This assumption reflects the fact that no more than one ordered DNA layer   can fit within the region $r<R_1$ if the inter-strand distance  exceeds $R_1$. The corresponding pitch condition $2\pi R_1 \tan \psi(R_1) = R_1$  leads to $\psi(R_1) = \arctan(\frac{1}{2\pi})$, and hence the  constant $M_1 = \ln \sin \psi(R_1)$.
 
  The boundary value problem defined by equations \eqref{hy-eqn}-\eqref{BC} has a unique solution satisfying $0< \psi <\frac{\pi}{2}$ \cite{liu2025knotted}. 
  We therefore conclude that the Oseen--Frank energy describing  DNA organization inside bacteriophage capsids admits   a unique minimizer of the director field.  Integration of this minimizing vector field  provides an estimate of the DNA spooling trajectory within the capsid.   

Next, we use the presented model  to estimate the ratio of  the Frank elasticity constants, $\alpha=\frac{K_3}{K_2}$, by fitting the results shown in Fig.\ref{fig:P4_layers}. The model input parameters are: the radius of the measured disordered phase, $R_1=10.00\ nm$; the  capsid  radius, $R_2=20.72\ nm$; and the radius of the outer ordered layer $r_1=17.69\ nm$. We performed a golden-section search to identify the value of $\alpha$ that best fit the radius of the remaining DNA layers, experimentally measured to be at $r_2=14.67 \ nm$ and $r_3=12.24 \ nm$. Our best fit was estimated  radii $\Tilde{r}_{2}=15.76\ nm$ and $\Tilde{r}_{3}=11.94 \ nm$, corresponding to  $\alpha=0.0064$. These  results suggest that, in bacteriophage P4, the twist elasticity constant is approximately two orders of magnitude larger than the bending constant.

\begin{figure}[htbp]
    \centering
\raisebox{-0.5\height}{\includegraphics[width=0.11\textwidth]{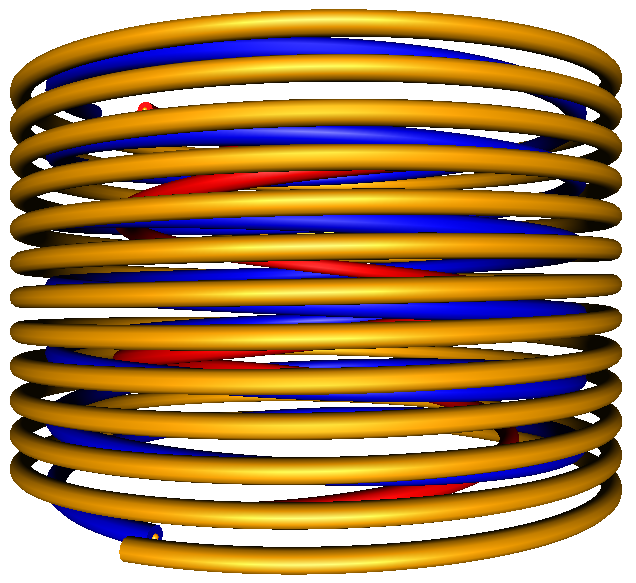}}
\raisebox{-0.5\height}{\includegraphics[width=0.11\textwidth]{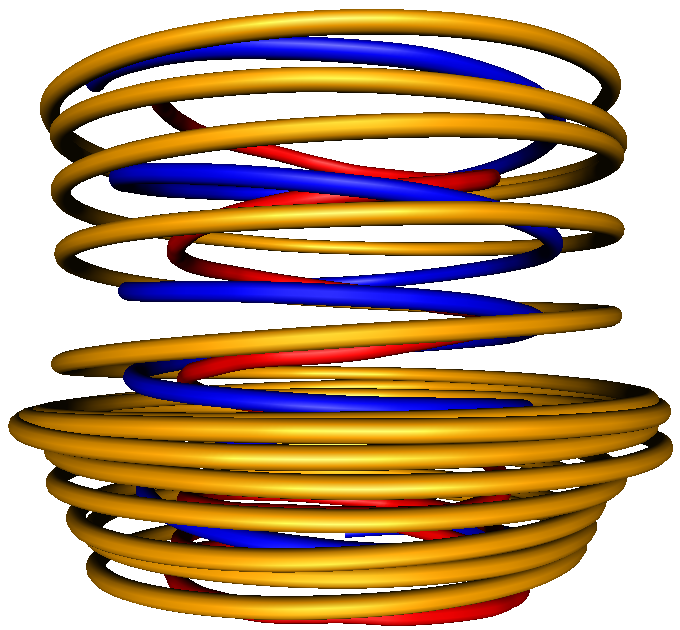}}
\raisebox{-0.5\height}{\includegraphics[width=0.11\textwidth]{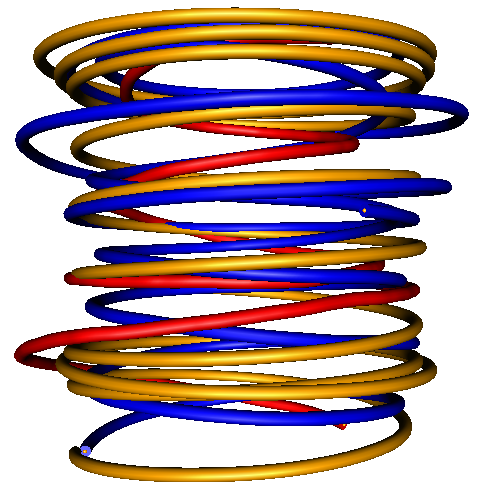}}
\raisebox{-0.5\height}{\includegraphics[width=0.11\textwidth]{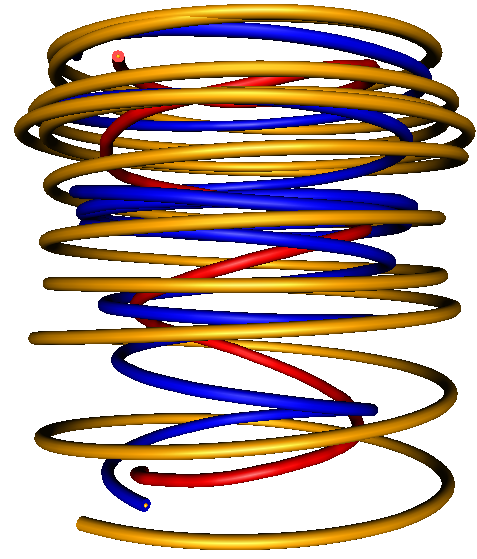}}
\par\vspace{1mm}
\text{ (a) \hspace{1.4cm} (b) \hspace{1.4cm} (c) \hspace{1.4cm} (d)}
    \caption{Multiple DNA simulations plotted with KnotPlot~\cite{Scharein2024}. Minimizer solution to the Euler-Lagrange equation (a). Simulated trajectories of a $3_1$ knot \protect\raisebox{-0.2ex}{\includegraphics[height=1em]{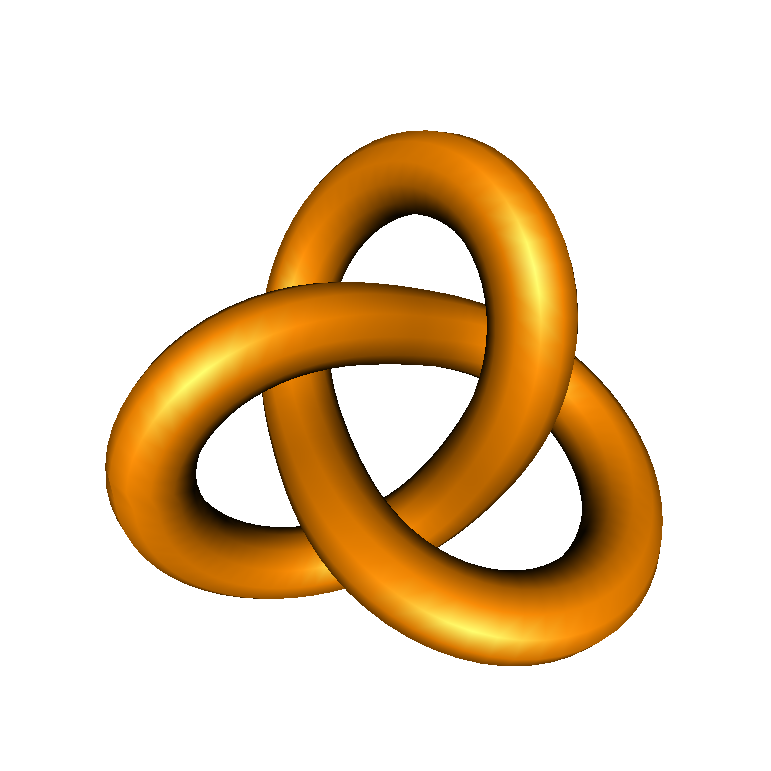}} (b), $4_1$ knot \protect\raisebox{-0.2ex}{\includegraphics[height=1em]{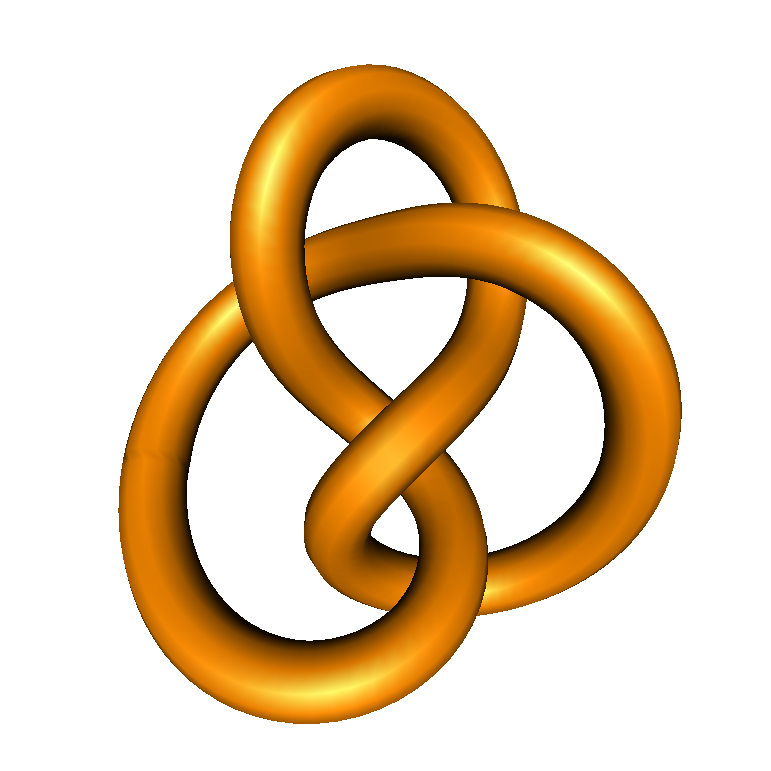}} (c) and $5_1$ knot \protect\raisebox{-0.2ex}{\includegraphics[height=1em]{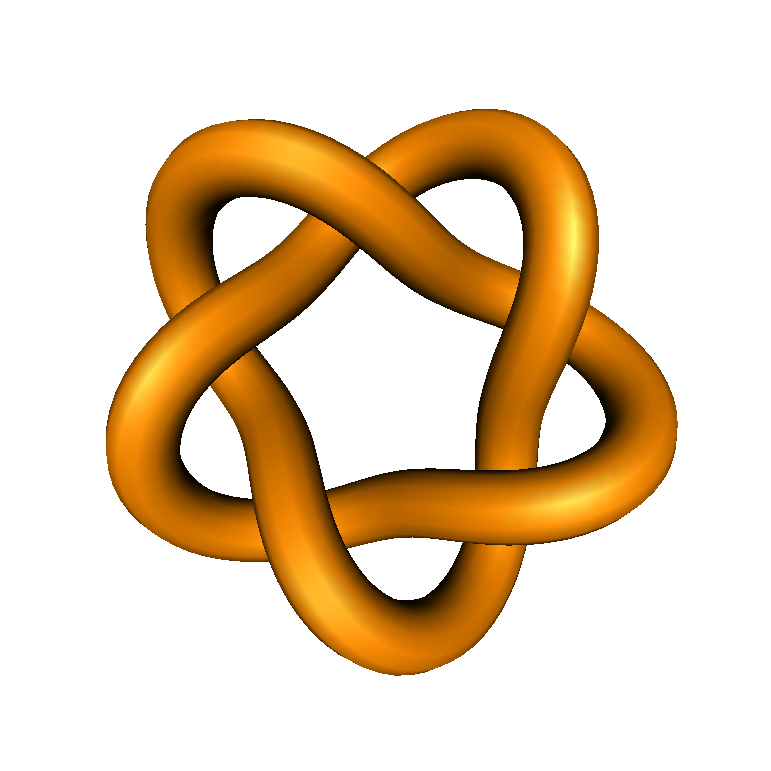}} (d)}
    \label{fig:randomized_dna}
\end{figure}

To validate this result we investigate whether the topological properties of the proposed minimizer are consistent with those observed in P4 experiments. 
DNA extracted from bacteriophage P4 capsids is knotted, and the observed knot distribution differs significantly from those obtained in random cyclization experiments of DNA molecules \cite{Liu1981,rybenkov1993probability}.  Furthermore, the knot distribution of P4 DNA features: a predominance of 
 trefoil $3_1$ and five crossing toroidal $5_1$ knots over other non-trivial knots, and by the low amounts of the four crossing $4_1$ and of the five crossing twist $5_2$ knots   \cite{Arsuaga2002b}. We therefore asked whether the knot distribution obtained from three-dimensional reconstructions of the P4 genome--using the value of $\alpha=0.0064$-- is consistent with the experimentally observed $P4$ knot distribution. 
 
As detailed in \cite{liu2025knotted}. The director field $\vec{n}_0 = (0 ,\cos \psi_0, \sin \psi_0)$, where $\psi_0$  the solution to the Euler--Lagrange equation, describes the vector field corresponding to the DNA trajectory. This deterministic  conformation does not account for entropic noise  and therefore cannot capture the knotted configurations observed in P4 capsids.
 To address this limitation, we introduce a stochastic term in the tangent direction $\vec{n}$ for each individual realization of the DNA trajectory, using a  Langevin-type equation.  Let 
\begin{equation}
    \vec{n} = (\sin \beta, \cos \beta \cos \psi, \cos \beta \sin \psi)
    \label{LangevinEquations}
\end{equation}
be the tangent vector after randomization. Then the trajectory of simulated DNA molecule is given by,
\begin{equation}
        \displaystyle \frac{dr}{ds} = \sin \beta, \quad
        \displaystyle \frac{d\theta}{ds} = \frac{\cos \beta \cos{\psi}}{r}, \quad
        \displaystyle \frac{dz}{ds} = \cos \beta \sin{\psi}.
    \label{Trajectory_Equations}
\end{equation}
The model depends on two angles: the helical angle $\psi$ and the  radial angle $\beta$. Perturbations of these two angles are given by the  following expressions:
\begin{eqnarray}
\begin{cases}
    d\beta = \sigma_\beta dB_\beta - \kappa_\beta  \beta ds,\\
    d\psi = \sigma_\psi dB_\psi - \kappa_\psi (\psi - \psi_0) ds
\end{cases}\label{LangevinNoise}
\end{eqnarray}

 The parameters  $\sigma_\beta$ and $\sigma_\psi$ represent the noise associated with Brownian motion and both scale as $\propto \sqrt{T}$, where  $T$ denotes temperature. The relaxation coefficients $\kappa_\beta$ and $\kappa_\psi$, which characterize the rate at which the tangent vector $\vec{n}$ relaxes to its equilibrium state $\vec{n}_0$  (i.e., the solution to the Euler-Lagrange equation), satisfy the asymptotic relation  $\kappa_\beta \propto K_1/r^3 \gg \kappa_\psi \propto (K_2+K_3)/r^3$. By combining equations \eqref{Trajectory_Equations} and \eqref{LangevinNoise}, we obtain a system of first-order differential equations whose solutions yield randomized DNA trajectories.

We computed the numerical solution using the Euler–Maruyama method \cite{Sauer2011numerical} for equation \eqref{LangevinNoise}, and the standard Euler method for equation \eqref{Trajectory_Equations} \cite{liu2025knotted}. 

\begin{figure}[htbp] 
\centering
\includegraphics[width=.95\linewidth]{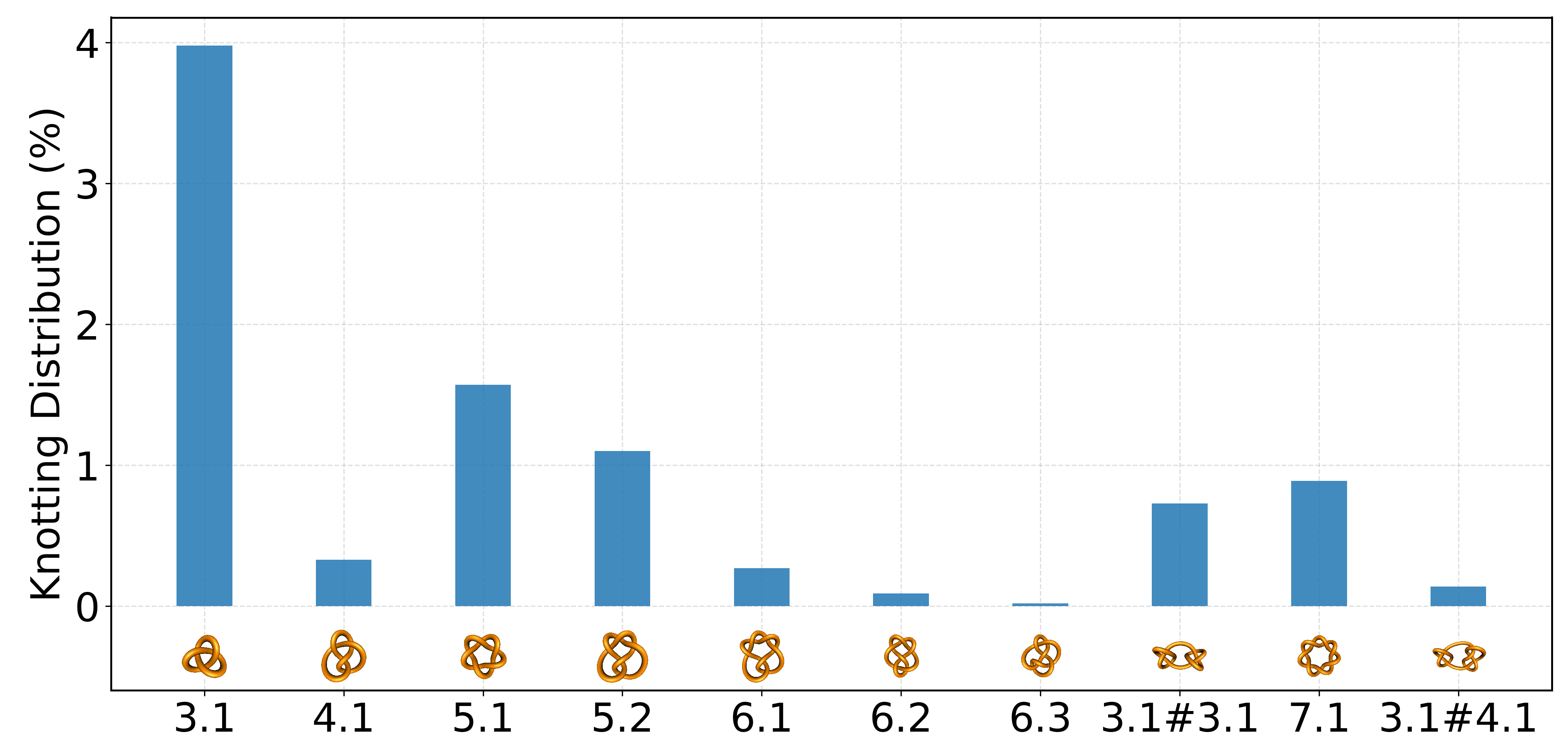} 
\caption{Knotting distribution obtained for 3 layers containing 2200 nm of DNA. The value of $\alpha=0.0064$.}\label{fig:P4_knots}
\end{figure}

To reflect experimental findings that show that approximately 90\% of DNA particles extracted from P4 phages are knotted \cite{Arsuaga2002a}, we selected parameter values that produced this proportion of knotted conformations. We then generated $10,000$ DNA trajectories and determined their knot types using the HOMFLY-PT polynomial as implemented in \cite{jenkins1989knot, marco2015libhomfly,Scharein2024}. The resulting knot distribution is shown in Fig.~\ref{fig:P4_knots}.

The distribution of knots in Fig. \ref{fig:P4_knots} differs markedly  from that produced by knots formed by   randomly embedded curves within spherical volumes \cite{Arsuaga2005,marenduzzo_computer_2008},  and it is in overall agreement with the experimental results reported in \cite{Arsuaga2005}. The simulated distribution exhibits  a predominance of  trefoil  knots $3_1$ and of the toroidal knot $5_1$ over the twist knot $5_2$ and the four-crossing knot $4_1$. 
The model overestimates the frequency  of the $5_2$ knot population, an effect also reported in all theoretical studies of P4 knotting \cite{Arsuaga2005, Arsuaga2008, Micheletti2008}. Interestingly, the simulated distribution reveals a significant population of the toroidal knot $7_1$ and populations of the connected sum of two trefoils $3_1\#3_1$ and $3_1\#4_1$ that are consistent with the data. 

{\it Discussion. \,} It is becoming increasingly evident that the liquid crystalline properties of DNA play a key role in the three dimensional organization of genomes. The development of mathematical models describing the liquid crystalline behavior in DNA or chromatin fibers has, however, lagged behind, in part due to the lack of experimental data needed to estimate the corresponding Frank elasticity constants. In this work, we introduce a combination of experimental and mathematical framework that enables us to estimate the relation between  these constants for bacteriophage DNA liquid crystals.

Our work extends existing biophysical and mathematical models that describe the behavior of DNA inside viral capsids. While earlier models captured some physical features of confined DNA, they did not fully develop a liquid crystalline model. The pioneering study by Klug and collaborators \cite{klug2003director} introduced key ingredients for modeling DNA within phage capsids; however, it provided only qualitative descriptions of the DNA trajectory. Subsequent works, such as \cite{walker2020fine,hiltner2018equilibrium}, employed the Oseen–Frank energy to investigate DNA organization in bacteriophage capsids. These studies, however, focused on bending contributions and hexagonal or cohesive interactions, but did not incorporate the cholesteric term.

In contrast, Marenduzzo et al. \cite{marenduzzo2009dna} introduced a cholesteric interaction into a standard molecular mechanics framework to simulate the knotted distributions observed in P4 capsids. More recently, full-atom simulations of entire bacteriophages have revealed DNA trajectories consistent with a liquid crystalline organization of the genome \cite{coshic2024structure}, though the specific role of the cholesteric term was not considered.

We find that the twist elastic constant $K_2$ is two orders of magnitude larger than the bending constant $K_3$ in bacteriophage capsids. This result  contrasts with previous estimates based on short segments of condensed DNA \cite{lucchetti2020elasticity} and theoretical predictions on polymers \cite{vroege1988induced}, which suggest  $\alpha=K_3/K_2>1$. Our findings  are consistent with experimental observations  of stable  DNA spooling structures of bacteriophages under a wide range of   ionic conditions.  

We have used Langevin  dynamics simulation framework to generate knotted trajectories consistent with the Oseen--Frank energy minimizer. The results of our simulations align with experimental observations and agree with the results reported in Monte-Carlo and molecular dynamics studies \cite{Arsuaga2008,marenduzzo2009dna}. The experimental data in \cite{Arsuaga2005} shows two populations for six and seven crossing knots. Our model predicts that for for the six crossing knots those two populations would correspond to the composite knot $3_1\#3_1$ and $6_1$, with the former having higher frequency. There are eight seven crossing knots and it is therefore very difficult to infer what populations may be present in the experimental data. Based on the frequency of knots with less number of crossings, we could conjecture that the two populations present are the  $7_1$ and $3_1\#4_1$. A key distinction between our approach and that of \cite{Arsuaga2008,marenduzzo2009dna}  is that our model is not guided by the desired knot distribution but instead by the spooling structure revealed in cryoEM data of bacteriophage P4,  a structural organization  that appears to be  overestimated in those works. 

Our results suggest important questions in the area of DNA modeling using liquid crystals and on the role of forces acting on the DNA inside viral capsids. With respect to the former, there is a strong motivation to further investigate the dependence of the Langevin dynamics parameters on temperature and ionic conditions  and to further extend this study to the de Gennes-Landau formulation for liquid crystals. Notably, comparing the Oseen-Frank energy with the  de Gennes-Landau free energy for the order tensor, reveals that  the Frank constants are expected to depend on concentration of the liquid crystal molecules. From a validation perspective, the de Gennes-landau framework offers an advantage: it naturally accommodates defects and knotted conformations, making it more suitable for comparison with experimental observations. With resect to the latter, one would like to determine the relationship between $K_2$ and ionic forces.  Qiu and colleagues reported that ionic interactions dominate the bending rigidity of DNA in bacteriophage lambda \cite{qiu2011salt}, therefore a promising direction is the incorporation of  explicit  electrostatic and Lennard Jones interactions, as implemented in \cite{liu2021ion}. 


\begin{figure}[!htbp] 
\begin{center}
\includegraphics[width=0.49\textwidth]{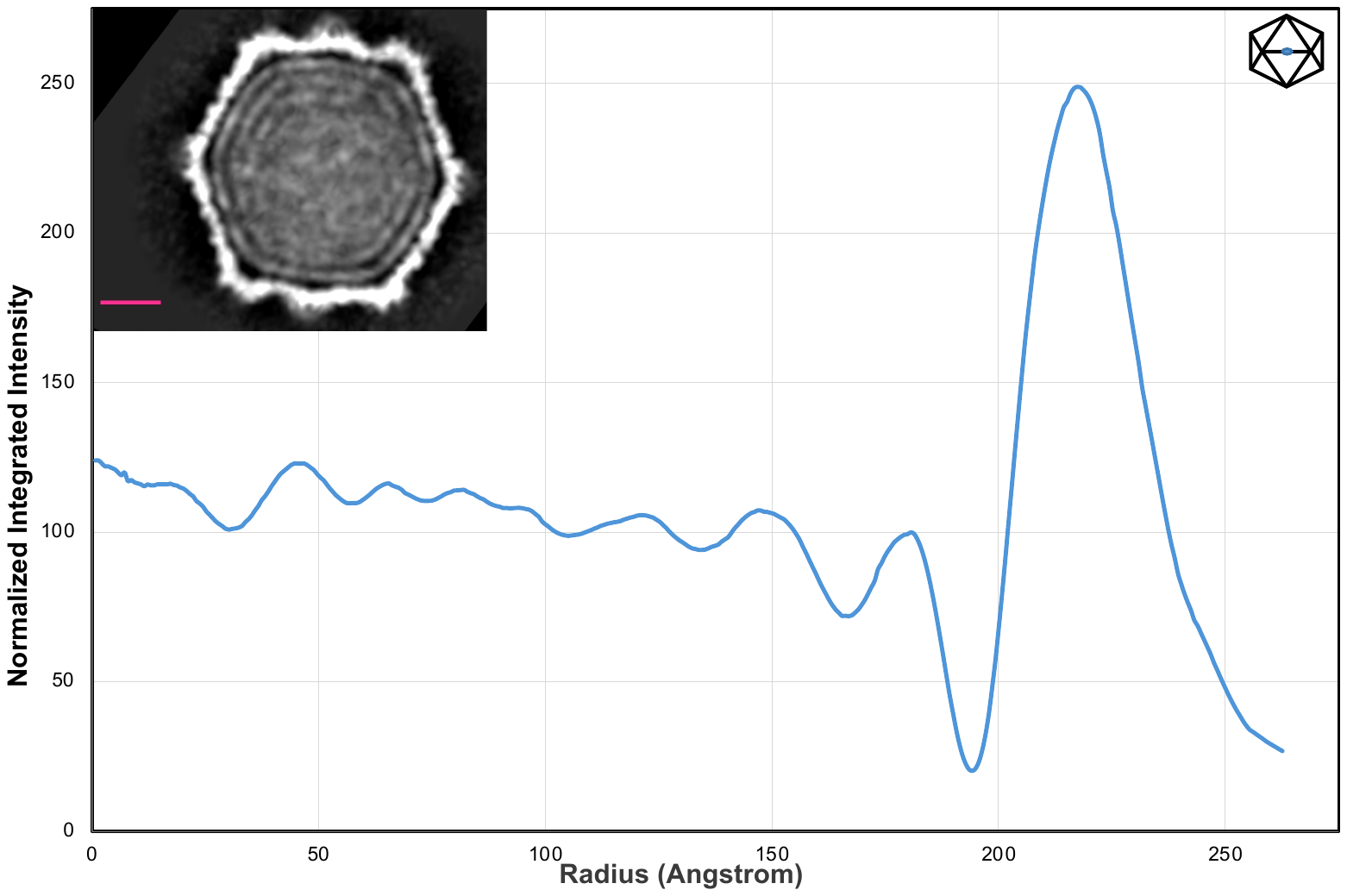}
\includegraphics[width=0.49\textwidth]{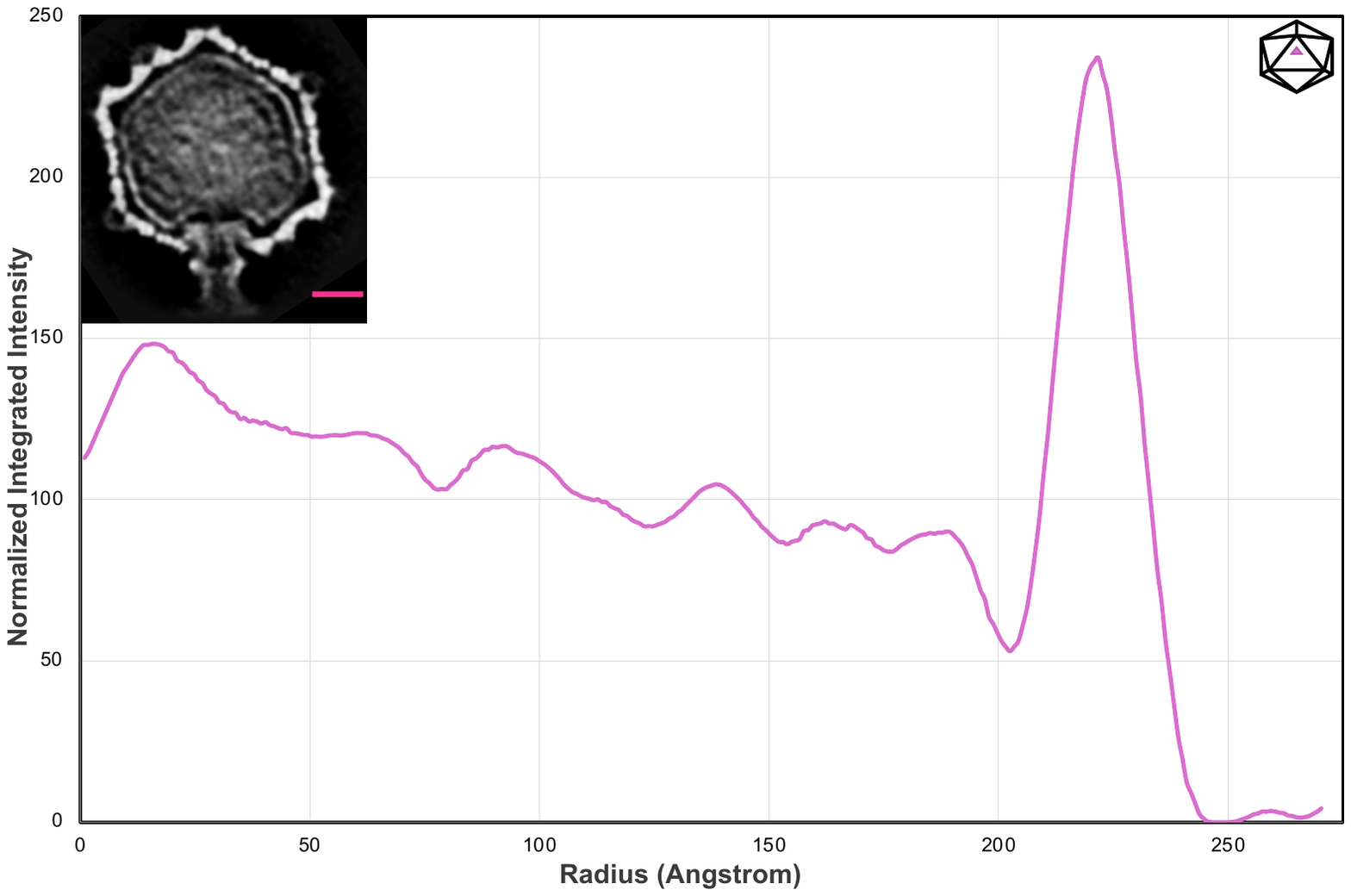}
\caption{Layer structure observed on P4 bacteriophages: Electron density maps as a function of the distance from the center of the capsid along a 2-fold and 3-fold axis. The inserts show the images of bacteriophage particles projected along a 2 and a 3 fold axis and a cartoon of an icosahedron highlighting these projections. }\label{fig:P4_layers_2fold}
\end{center}
\end{figure}

\section*{Appendix} 

The following three steps follow the first item in the {\it Materials and Methods} section of the main paper.
We have placed them separately since they describe standard techniques in the subject.

\noindent {\bf Cryo-EM sample preparation} Aliquots ($<5$ ul) of highly concentrated ($>10^{12}$ plaque-forming units per milliliter) purified viral stock was applied to glow-discharged holey carbon grids and then cryo-plunged into liquid ethane after a brief incubation and blotting period \cite{adrian1984cryo}. 

\noindent  {\bf Cryo-EM data collection} 
Images were acquired using a 200kv Thermo Scientific$^{\text{TM}}$ Glacios$^{\text{TM}}$ Cryo Transmission Electron Microscope equipped with a K3 Direct Electron Detector and X-FEG optics using the automation software SerialEM. Microscope parameters applied to acquired data set were the following: spherical aberration 2.7 $mm$, total electron dose  0.80 e- per$>\text{\AA}^{2}$  per frame (75 frames/ micrograph), super resolution pixel size of 0.44 \AA per pixel , and defocus between 0.4 $\mu m$ and 4.0 $\mu m$. 

\noindent {\bf Cryo-EM reconstruction} 
Image analysis, particle selection, and 3D structure formation were all completed using    cryoSPARC \cite{punjani2017cryosparc}. Microscope-derived image anomalies in the data were Contrast Transfer Function (CTF) corrected by manual inspection. 7,321 selected particles using an initial box size of 1660 pixels which were Fourier cropped to 830. Following 2D classification, selected classed were used to construct an initial model with I1 symmetry imposed. The model was then refined without symmetry imposed. The reconstruction obtained has a resolution of 24.88 \AA. The central cross-section was generated using ChimeraX \cite{pettersen2021ucsf}. Fiji software \cite{schindelin2012fiji} was used to preform the radial intensity profile using the Radial Profile pluggin.

{\bf Number of layers in bacteriophage P4}
Fig \ref{fig:P4_layers_2fold} shows the number of layers when projecting along a two and a three fold axis. 

\begin{acknowledgments}
P.L. was partially supported by NSF grant DMS-2318053. T.C., Z.W. and J.A. were partially supported by NSF grant DMS-2318052 and DMS-1817156. MC.C. was partially supported by NSF grant DMS-2318051. M.V. was partially supported by NSF grant DMS-2054347 and DMS-1817156. 
\end{acknowledgments}

\bibliography{gel, pei, chromonic}

\end{document}